\PassOptionsToPackage{table}{xcolor}
\documentclass[conference]{IEEEtran}
\IEEEoverridecommandlockouts
\usepackage{cite}
\usepackage{amsmath,amssymb,amsfonts,balance}
\usepackage{graphicx}
\usepackage{textcomp}
\usepackage{xcolor}
\usepackage{algorithm}
\usepackage{algpseudocode}
\usepackage{mathtools}
\usepackage[numbers,sort&compress]{natbib}
\usepackage{multirow}
\usepackage{tabularx, hhline}
\newcommand{\ie}{{i.e.,~}}
\newcommand{\eg}{{e.g.,~}}

\newcommand{\etal}{{et~al.}}

\def\BibTeX{{\rm B\kern-.05em{\sc i\kern-.025em b}\kern-.08em
    T\kern-.1667em\lower.7ex\hbox{E}\kern-.125emX}}
\begin{document}
\title{CSUM: A Novel Mechanism for Updating CubeSat while Preserving Authenticity and Integrity}
\author{
	Ankit Gangwal, Aashish Paliwal\\
	International Institute of Information Technology Hyderabad, India\\
	\textit{gangwal@iiit.ac.in, aashish.paliwal@research.iiit.ac.in}
}
\maketitle

\begin{abstract}
The recent rise of CubeSat has revolutionized global space explorations, as it offers cost-effective solutions for low-orbit space applications (including climate monitoring, weather measurements, communications, and earth observation). A salient feature of CubeSat is that applications currently on-boarded can either be updated or entirely replaced by new applications via software updates, which allows reusing in-orbit hardware, reduces space debris, and saves cost as well as time. Securing software updates employing traditional methods (e.g., encryption) remains impractical mainly due to the low-resource capabilities of CubeSat. Therefore, the security of software updates for CubeSats remains a critical issue.
\par
In this paper, we propose CubeSat Update Mechanism~(CSUM), a lightweight scheme to provide integrity, authentication, and data freshness guarantees for software update broadcasts to CubeSats using a hash chain. We empirically evaluate our proof of concept implementation to demonstrate the feasibility and effectiveness of our approach. CSUM can validate 50,000 consecutive updates successfully in less than a second. We also perform a comparative analysis of different cryptographic primitives. Our empirical evaluations show that the hash-based approach is at least 61$\times$ faster than the conventional mechanisms, even in resource-constrained environments. Finally, we discuss the limitations, challenges, and potential future research directions for CubeSat software update procedures.
\end{abstract}


\section{Introduction}
Space exploration and deployment of satellites have become an indispensable part of modern science. These satellites serve various purposes, including navigation~\cite{lechner2000global}, weather forecast~\cite{kalsi2002satellite}, and data collection~\cite{antonini2005satellite}. In the past few years, CubeSat~(CS) - a novel type of miniature satellite - has attracted significant interest from researchers in academia as well as industry~\cite{swartwout2013first}. Each CS unit measures $10 \times 10 \times 10$ cubic centimeters~($10\ cm^3$). The rise of CS has significantly reduced the cost associated with satellite deployment, consequently democratizing the field. In particular, the reduction in satellite deployment cost has enabled small to mid-sized companies, universities, research institutes, and even developing nations~\cite{woellert2011cubesats} to participate in space expeditions.
\par
While CS provides a cost-effective means for entering the satellite playground, it also introduces a new array of challenges that do not apply to traditional satellites. The fundamental reason for such unprecedented challenges is the limited availability of resources~(including computing power, memory, and communication bandwidth) in CS. One such challenge within resource-constrained environments is the software update process~\cite{bellissimo2006secure}, and CSs are no different. In fact, it is even more crucial to secure the software update process for CSs as they are mission-critical, where even a minute error can lead to mission failure or the complete loss of a satellite. Furthermore, it is essential for CS's software to be modular and reusable~\cite{eshaq2023flight}.
\par
Malicious software updates can lead to unauthorized control, data manipulation, and communication interference, which can lead to severe consequences~(e.g., increased risk for mission failure, lack of data integrity, even complete loss of control over CS). Moreover, unauthorized access to a satellite poses a significant risk not just to its owners, but to the broader space environment as well. As a representative example, an attacker taking control of a satellite and activating its thrusters could lead to the Kessler Syndrome~\cite{kessler2010kessler}. Kessler Syndrome is a scenario where debris from one satellite collision spreads and hits other satellites, creating more debris in a domino effect. Such a chain reaction could potentially block access to space for decades, as observed in different simulations~\cite{pavur2019cyber, drmola2018kessler}. CS commonly utilizes amateur radio frequencies for communication~(i.e., UHF/VHF~\cite{molina2023cubedate}), resulting in relatively low data rates~(typically ranging from 9.6~Kbps to 100~Kbps). For instance, the daily throughput of ThingSat~\cite{csug-thingsat} is approximately 1500~KB~\cite{molina2023cubedate}. Furthermore, utilizing standard cryptographic schemes to handle malicious updates is impractical since cryptographic operations tend to be expensive in terms of computational resources; making them unsuitable for CS's resource-constrained environment.
\par
The on-board software is one of the most critical components of any space mission. It encompasses the core capabilities of the space system, ranging from daily activities within the system~(like navigation, communication, and energy management) to more specialized tasks~(like data collection and processing). The effectiveness of a space system relies heavily on the correct functioning of on-board hardware and software. Post-launch, satellites may face unexpected events and operating conditions. While some issues can be mitigated via exception handling~(if anticipated during initial software development), others may be fixed through software updates. 
\par
Over-the-Air~(OTA) updates have overwhelmingly affected various user-centric domains, e.g., smartphones, smart TVs, automobiles, and IoT devices. OTA enables manufacturers to deliver software updates remotely, eliminating the need for physical access to a device. For the end users, OTA updates offer a convenient way to receive the latest functionality without visiting the service center. For manufacturers, it provides a way to improve their products' value by delivering new features, bug fixes, and security patches on time. The importance of OTA updates extends beyond convenience. OTA updates are critical for the lifecycle management of modern devices to ensure they remain efficient, secure, and up-to-date with minor user intervention. 
\par
Nilchiani~\cite{nilchiani2009valuing} argues that the capability of remote modification and enhancement of on-board software is crucial for space systems due to the uncertain nature of space. This capability can help preserve system functionality, adapt to emerging requirements, or improve performance without needing physical upgrades (which are often impossible after launch). In-orbit software updates are now standard practice for small satellites~\cite{garrido1998minisat01}. Even nano-satellites~(\eg PlanetLabs Dove nanosatellite constellation~\cite{marshall2013planet}) acknowledge support for in-orbit firmware updates without disclosing the details.
\par
In this paper, we introduce CSUM, a lightweight scheme designed to enhance the security of software updates in CS. CSUM is designed to take into account the constrained hardware capabilities of CS and the limited communication bandwidth between Ground Station (GS) and CS. It aims to preserve software updates' integrity, authenticity, and freshness, thereby safeguarding CS from adversaries attempting to tamper with in-transmission software updates. Our approach advocates for using lightweight hash functions because a single public-key computation is roughly equivalent to hundreds of hash computations in processing time~\cite{park2018one}.
\par
The major contributions of our paper are as follows:
\begin{enumerate}
    \item We propose CSUM, a novel and lightweight scheme that utilizes hash chains to ensure authentication, integrity, and freshness for CS software update broadcasts.
    \item We validate the effectiveness of CSUM via empirical evaluations of its proof of concept implementation. Our results show that CSUM can validate 50,000 consecutive updates in just 0.81 seconds.
    \item Furthermore, we perform a comparative analysis of different cryptographic primitives. Our analyses show that the hash-based approach outperforms the traditional mechanisms even in resource-constrained environments. In particular, encryption, decryption, signing, and signature verification operations are over 155, 126, 64, and 61 times slower than the hash-based approach, respectively.
\end{enumerate}
\textit{Organization:} Section~\ref{section:background} provides an overview of the related research works and background knowledge. We discuss our system and adversary model along with security requirements in Section~\ref{section:system_adversary_model}, elaborate CSUM in Section~\ref{section:proposed_scheme}, and discuss our results in Section~\ref{section:evaluation}. We present CSUM's security analysis in Section~\ref{section:security_analysis}. Section~\ref{section:discussion} highlights the potential limitations of our proposed solution. Finally, Section~\ref{section:conclusion} concludes the paper and highlights the possible future directions for CS software update procedure.

\section{Background}
\label{section:background}
Section~\ref{subsection:related_work} delves into the advancements in securing space systems, particularly through OTA software updates, and highlights the challenges in securing software updates in resource-constrained environments. Section~\ref{subsection:prerequisite} introduces cryptographic fundamentals essential for our proposed scheme.

\subsection{Related works}
\label{subsection:related_work}
The security of space-based assets (like CS) has not been extensively researched as their terrestrial counterparts (like connected vehicles), especially concerning OTA software updates. Recently, researchers have been looking into CS security, focusing on the unique challenges posed by the limited resources available on such platforms. Halder~\etal~\cite{halder2020secure} underlines the significance of OTA updates and categorizes existing OTA update techniques for connected vehicles. Various secure update techniques for connected vehicles have been developed to address integrity, authenticity, and confidentiality. These techniques include Uptane~\cite{karthik2016uptane}, a secure software repository framework that enhances compromise resilience by distributing responsibilities across distinct roles. Blockchain-based schemes~\cite{baza2019blockchain} eliminate cloud involvement and utilize smart contracts to ensure update integrity and authenticity, while hash function-based protocols~\cite{nilsson2008secure} safeguard the transmission integrity of software updates. Frameworks, such as SecUp~\cite{steger2017efficient}, utilize a combination of symmetric and asymmetric key cryptography to ensure secure and efficient OTA updates. Moreover, hardware-based solutions leveraging Hardware Security Modules (HSM)~\cite{idrees2011secure} and Trusted Platform Modules (TPM)~\cite{petri2016evaluation} provide robust security at hardware-level. 
\par
Souza~\etal~\cite{de2022flight} emphasize the difficulties in implementing secure software updates due to these resource constraints, proposing a multi-layered mission software approach for CS. Similarly, Bellissimo~\etal~\cite{bellissimo2006secure} highlights the challenges in securing software updates in deployed systems, complexities in safeguarding against known attacks, and the challenges of applying secure content distribution methods in resource-limited devices. Willbold~\etal~\cite{willbold2023space} offers a comprehensive threat taxonomy against satellite firmware, including an analysis of real-world satellite firmware security issues and a survey among professional satellite developers to shed light on the satellite security landscape. 
\par
The authors in~\cite{fitzsimmons2012reliable, sunter2016firmware, molina2023cubedate} discuss different firmware update mechanisms for an in-orbit CS. Sünter~\etal~\cite{sunter2016firmware} explore firmware updating systems for nano-satellites using the ESTCube-1 mission~\cite{slavinskis2015estcube} to compare four distinct update procedures, highlighting their implementation complexities and error recovery mechanisms for in-orbit satellite software management. Fitzsimmons~\cite{fitzsimmons2012reliable} improves CS robustness through a software update mechanism, focusing on update usability, validation, and system recovery to extend mission capabilities and safeguard against operational anomalies. The author utilizes MD5 for integrity, but the scheme does not address authentication. Bezem and Fjellby~\cite{bezem2012authenticated} tackle the absence of inherent security features in satellite communications by enhancing the CS Space Protocol with HMAC and sequence numbers to prevent replay attacks. However, apart from focusing exclusively on replay attacks, the scheme also requires secure infrastructure, which is comparatively costly. Molina~\etal~\cite{molina2023cubedate} addresses secure software updates for multi-tenant CS through Cubedate, a framework designed for continuous software deployment to orbiting CS. Despite its innovative approach, the framework relies on digital signatures and encryption, raising concerns about computational feasibility in the constrained environment, alongside its dependence on a single trust anchor. Challa~\etal~\cite{challa2012cubesec} proposed CubeSec and GndSec, lightweight security solutions for CS communications, acknowledging the need for efficient security measures in resource-constrained environments. However, the security of their approach relies heavily on pre-shared keys as they utilize symmetric encryption.
\par
In traditional pre-shared key systems, all security properties are lost once the key (or trust anchor) is compromised. In our approach, even if a CS is compromised, adversaries can only validate tokens and read the next one when it comes; they cannot create the next key due to the pre-image resistance of cryptographic hash functions. This significantly limits the damage an adversary can do as long as they can only read the trust anchor and not replace it.
\par
To summarize, using
standard cryptographic primitives~(i.e., encryption in CubeSec~\cite{challa2012cubesec}, CubeDate~\cite{molina2023cubedate}, and GndSec~\cite{challa2012cubesec} as well as digital signatures in CubeDate~\cite{molina2023cubedate} and NUTS~\cite{bezem2012authenticated}) make state-of-the-art CS security solutions computationally expensive. \textbf{CSUM overcomes computational overheads by limiting the use of public-key operations to just one instance~(i.e., to create a one-time trust anchor) and utilizing lightweight hash function in subsequent operations.} 

\subsection{Prerequisites}
\label{subsection:prerequisite}
    
\subsubsection{Cryptographic hash function}
A hash function transforms a message of any length into a random-looking fixed-size string. For a hash function to be considered a cryptographic hash function, it must contain three properties:
\begin{itemize}
    \item Pre-image resistance: Given a hash function output $H$, it should be computationally infeasible to find input value $m$ such that $h(m) = H$.
    \item Second pre-image resistance: Given an input value $m_1$, it should be computationally infeasible to find a different input value $m_2$ such that \(h(m_1) = h(m_2), \text{where } m_1 \neq m_2\).
    \item Collision resistance: It should be computationally infeasible to find two distinct inputs $m_1$ and $m_2$ such that \(h(m_1) = h(m_2)\).
\end{itemize}
    
\subsubsection{Hash chain}
\begin{figure}[!ht]
    \centering
    \includegraphics[width=\columnwidth]{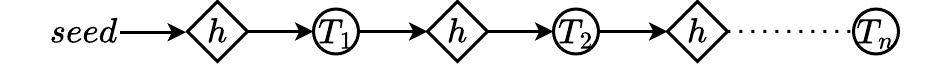}
    \caption{Representation of a hash chain starting from seed to $T_n$.}
    \label{fig:hash_chain}
\end{figure}

Hash chain utilizes a cryptographic hash function to create a linked chain of hashes. A hash chain is constructed by recursively applying a cryptographic hash function to a randomly generated seed as shown in \figurename~\ref{fig:hash_chain}. Hash chains inherently inherit the properties of cryptographic hash functions while also exhibiting their own unique characteristics due to their iterative structure.

\section{System architecture and threat model}
\label{section:system_adversary_model}
In this section, we outline the system model~(cf. Section~\ref{subsection:system_model}), describe the adversary model for CSUM~(cf. Section~\ref{subsection:adversary_model}), and specified the security requirements for CS~(cf. Section.~\ref{subsection:security_goals}).
\subsection{System model}
\label{subsection:system_model}
CSUM focuses on ensuring the secure delivery of Software Update Package~($SUP$) from GS to CS while considering the inherent vulnerabilities of the communication systems linking them. Our system model comprises three primary entities: administrator, GS, and CS. The administrator plays a pivotal role in the initial setup by generating a seed to create a hash chain, which acts as a trust anchor and lays the foundation for secure communication. The administrator is also responsible for creating $SUP$s and corresponding Transmission Token~($TT$), which are essential for authorizing updates. GS is responsible for transmitting $SUP$s and its corresponding $TT$s received from the administrator to CS. CS is tasked with the reception and verification of $SUP$ and their accompanying $TT$s, ensuring the security of the deployment process. CS is equipped to perform critical cryptographic operations despite its limited resources.
\par
The communication link between GS and CS is inherently insecure but reliable and susceptible to threats such as replay attacks, message alteration, and injection. Despite these challenges, both the administrator and CS are considered secure entities with secure internal storage capabilities to resist direct attacks. Operating within a constrained environment, CS must efficiently manage its limited bandwidth, processing capabilities, and storage, highlighting the need for streamlined and lightweight security solutions. A reliable transfer protocol, like Saratoga protocol~\cite{wood2007saratoga}, can mitigate issues related to packet losses and propagation delay. Our proposed scheme is agnostic to transport mechanisms as long as we transmit both $SUP$ and the corresponding $TT$.
\par
\begin{figure}[!ht]
    \centering
    \includegraphics[width=\columnwidth]{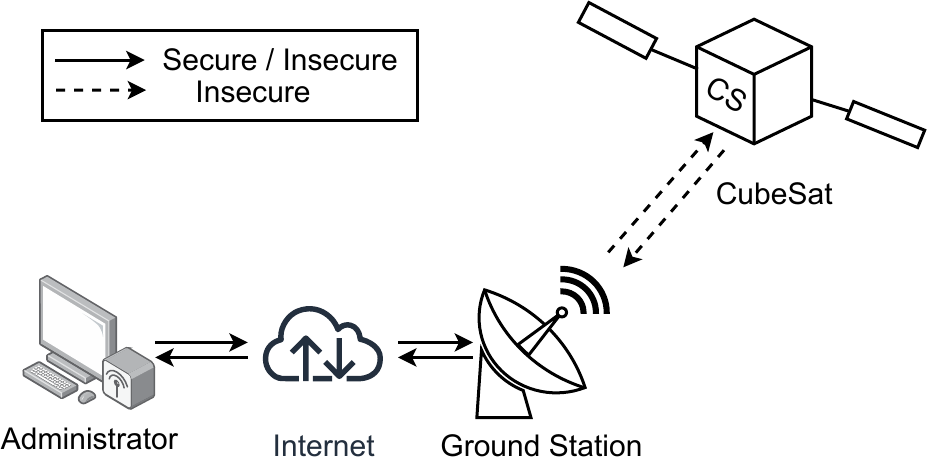}
    \caption{A simplified communication architecture of CS.}
    \label{fig:communication}
\end{figure}

 
\figurename~\ref{fig:communication} illustrates a communication system where the administrator interacts with the Ground Station over the Internet, with the potential for both secure and insecure data transmission. However, the communication between the Ground Station and the CS is insecure. 

\subsection{Adversary Model}
\label{subsection:adversary_model}
Our adversary model considers a highly capable adversary with computational resources surpassing those typically available in a CS environment. The adversary operates within Probabilistic Polynomial Time (PPT) and targets vulnerabilities primarily within the communication framework connecting GS and CS. We assume that the trusted components in the overall CS infrastructure, including the administrator, CS, and GS, remain secure throughout the system's operational life.
\par
The adversary possesses a diverse array of attack capabilities to compromise software updates between GS and CS. These capabilities include passive eavesdropping on the communication channel, active interception and data manipulation to facilitate Man-in-the-Middle attacks, executing replay attacks by re-transmitting legitimate $SUP$, injecting malicious $SUP$, and overwhelming CS~(and its verification capacity) by flooding CS with fake $SUP$ and $TT$.
\par
The adversary's objectives include disrupting genuine communication, deceiving CS into accepting malicious $SUP$s, extracting sensitive information from transmitted messages, and wasting CS resources. Despite these capabilities, the adversary is bound by PPT, which restricts its ability to break strong cryptographic systems within a reasonable time frame. Additionally, the adversary is incapable of altering the initially trusted data placed in a CS~(e.g., trust anchor).

\subsection{Security Goals}
\label{subsection:security_goals}
The primary objective of CSUM is to secure the transmission of $SUP$s from GS to CS against threats on insecure channels. We aim to achieve the following critical security goals~\cite{boyd2020protocols}:
\begin{enumerate}
    \item \textit{Authentication:} To ensure that $SUP$s originate from a verified source, CSUM incorporates the use of authentication tokens. This mechanism ensures that only authorized entities can initiate software updates.
    \item \textit{Integrity:} CSUM is designed to detect any unauthorized modifications in $SUP$s received by CS. Thereby safeguarding the integrity of the transmitted data.
    \item \textit{Data freshness:} It is crucial that CS receives the most recent $SUP$. CSUM mitigates the risk of replayed attacks by adversaries, ensuring that only the latest valid $SUP$ are installed.
\end{enumerate}

\section{Proposed Scheme}
\label{section:proposed_scheme}
CS is designed to operate under limited resources, e.g., extremely low bandwidth, constrained memory, and minimal processing power. Formulating effective security strategies within these constraints is non-trivial and presents a significant challenge. These constraints force us to move away from traditional cryptographic mechanisms that are robust but resource-intensive. Thus, such mechanisms are impractical for low-resource environments, like CS. In contrast, cryptographic hash functions emerge as a better solution due to their significantly lower computational overheads~\cite{park2018one}. CSUM utilizes a lightweight hash function to ensure the authentication and integrity of $SUP$s sent from GS to CS.
\par
CSUM aims to minimize reliance on resource-intensive public-key operations and replace them with efficient hash operations instead. This approach holds significant advantages for CS due to the lightweight nature of hash functions compared to encryption and signature schemes. The proposed scheme aims to achieve authentication, integrity, and freshness with constant network overhead~(\ie output length of a single hash function).
\par
Now we elucidate different phases~(cf. Section~\ref{subsection:phases}) and  operations~(cf. Section~\ref{subsection:operations}) in CSUM. We also present a sample execution of CSUM in Section~\ref{subsection:run}.

\subsection{Phases}
\label{subsection:phases}
CSUM involves setup, key encapsulation, and authentication and integrity phases.
We utilize several symbols as defined in \tablename~\ref{tab:symbol_def} while explaining CSUM.
 \begin{table}[H] 
	\centering
	\caption{Symbol definitions.}
	\begin{tabular}{ll}
		\hline
		\textbf{Symbol} & \textbf{Description} \\
		\hline
            $TA$          & Trust Anchor \\
		$TT$          & Transmission Token \\
		$DT$          & Derived Token (from transmission) \\
		$PT$		  & Partial Token \\
		$AT$	      & Authentication Token \\
            $AT_{curr}$	  & Current Authentication Token \\
            $AT_{prev}$   & Previous Authentication Token \\
		$h(x)$        & Hash function applied on $x$ \\
		$h^n(x)$      & Hash function applied on $x$ $n$ times \\
		\hline
	\end{tabular}
 \label{tab:symbol_def}
\end{table}

\subsubsection{Setup phase}
\label{subsection:setup_phase}
The setup phase initializes the system, where the administrator employs a hardware random number generator~\cite{davies2000hardware} to produce a random seed. The generated seed undergoes iterative hashing to form a hash chain~\cite{lamport1981password}. The resulting tip, $h^n(seed)$, is stored in CS memory as a Trust Anchor ($TA$) for subsequent operations.\\
Prior to launch, storing $TA$ in CS is straightforward. Post-launch introduces considerations for secure transmission. Confidentiality is ensured through pre-existing mechanisms supporting confidentiality, allowing direct transfer of $TA$ without additional security layers. Encryption becomes necessary only when the default mechanism lacks confidentiality support despite the additional overhead.

\subsubsection{Key encapsulation phase}
\label{subsection:key_encapsulation_phase}
The key encapsulation phase conceals a one-time Authentication Token~(AT) without utilizing encryption techniques. Our key encapsulation involves a bitwise XOR operation ($\oplus$) between the current $AT$~(\ie $AT_{curr}$) and the result of applying a cryptographic hash function ($h$) to the concatenation of $SUP$ and the previous $AT$~(\ie $AT_{prev}$) as shown in Eq.~\eqref{eq:transmission_token}. 
\begin{equation}
    \label{eq:transmission_token}
    TT \coloneqq AT_{curr} \oplus h(SUP \mathbin\Vert AT_{prev})
\end{equation}
\par
\begin{figure}[ht!]
    \centering
    \includegraphics[width=0.95\columnwidth]{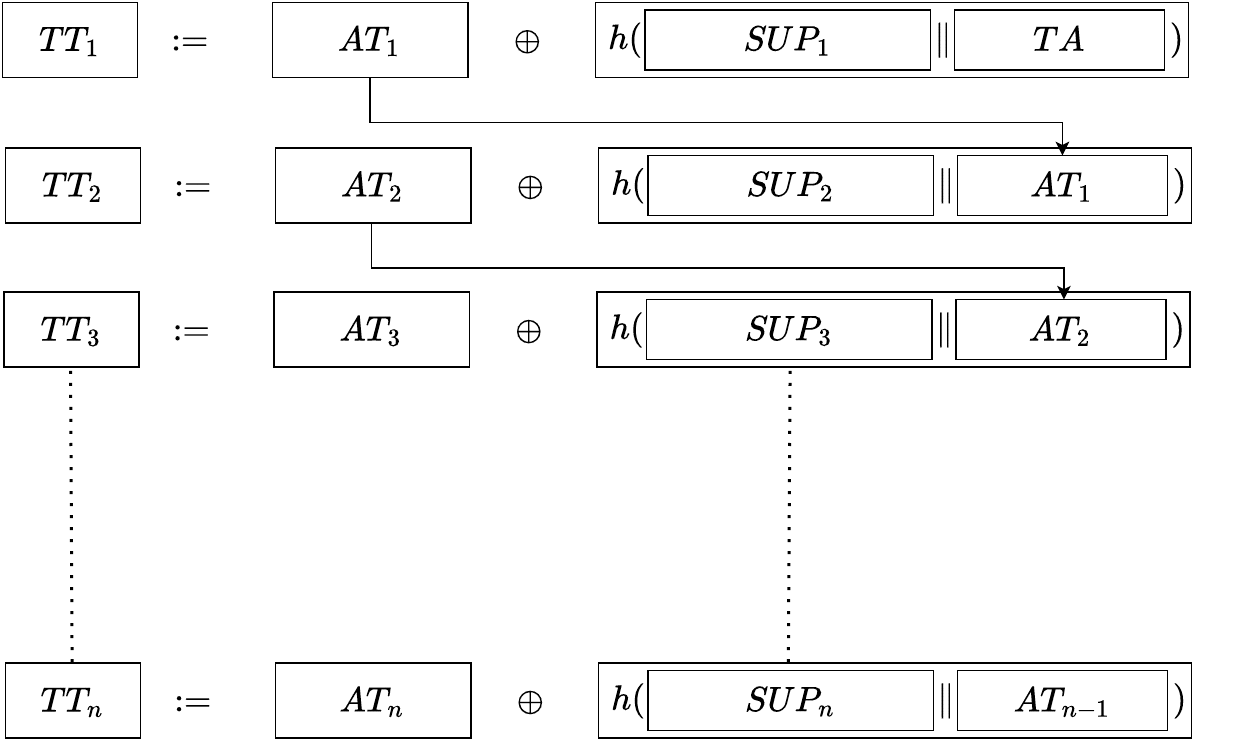}
    \caption{Protocol describing secure software update procedure.}
    \label{fig:protocol}
\end{figure}

In simpler terms, we combine $AT_{curr}$ with a hashed value derived from the concatenation of $SUP$ and $AT_{prev}$ to generate $TT$. \figurename~\ref{fig:protocol} shows interconnection between subsequent $TT$s.

\subsubsection{Authentication and integrity phase}
\label{subsection:authentication_integrity_phase}
The authentication and integrity phase is crucial in ensuring the secure transmission of $SUP$s from GS to CS. This phase links $TT$ to $AT_{curr}$, $SUP$, and $AT_{prev}$ (previous state of hash chain) while confirming the validity of $AT_{curr}$. CSUM ensures integrity by associating the hash of $SUP$ to $TT$ as shown in Eq.~\eqref{eq:transmission_token}. Any attempt by an adversary to alter $SUP$ results in CS failing to extract the correct $AT_{curr}$ from $TT$, leading to the rejection of $SUP$. $AT_{curr}$ verification requires a single hash operation, as shown in Eq.~\eqref{eq:auth}.
\begin{equation}\label{eq:auth}
    h(AT_{curr}) = AT_{prev}
\end{equation}

Overall, the proposed scheme leverages hash functions to ensure authentication and integrity in a resource-constrained environment, making it suitable for CS with limited bandwidth, processing power, and memory. The hash chain and $TT$ mechanisms provide a lightweight yet effective approach to secure software updates in the presence of potential adversaries and insecure communication channels.

\subsection{Operations}
\label{subsection:operations}
\begin{figure}[!ht]
    \centering
    \includegraphics[width=0.99\columnwidth]{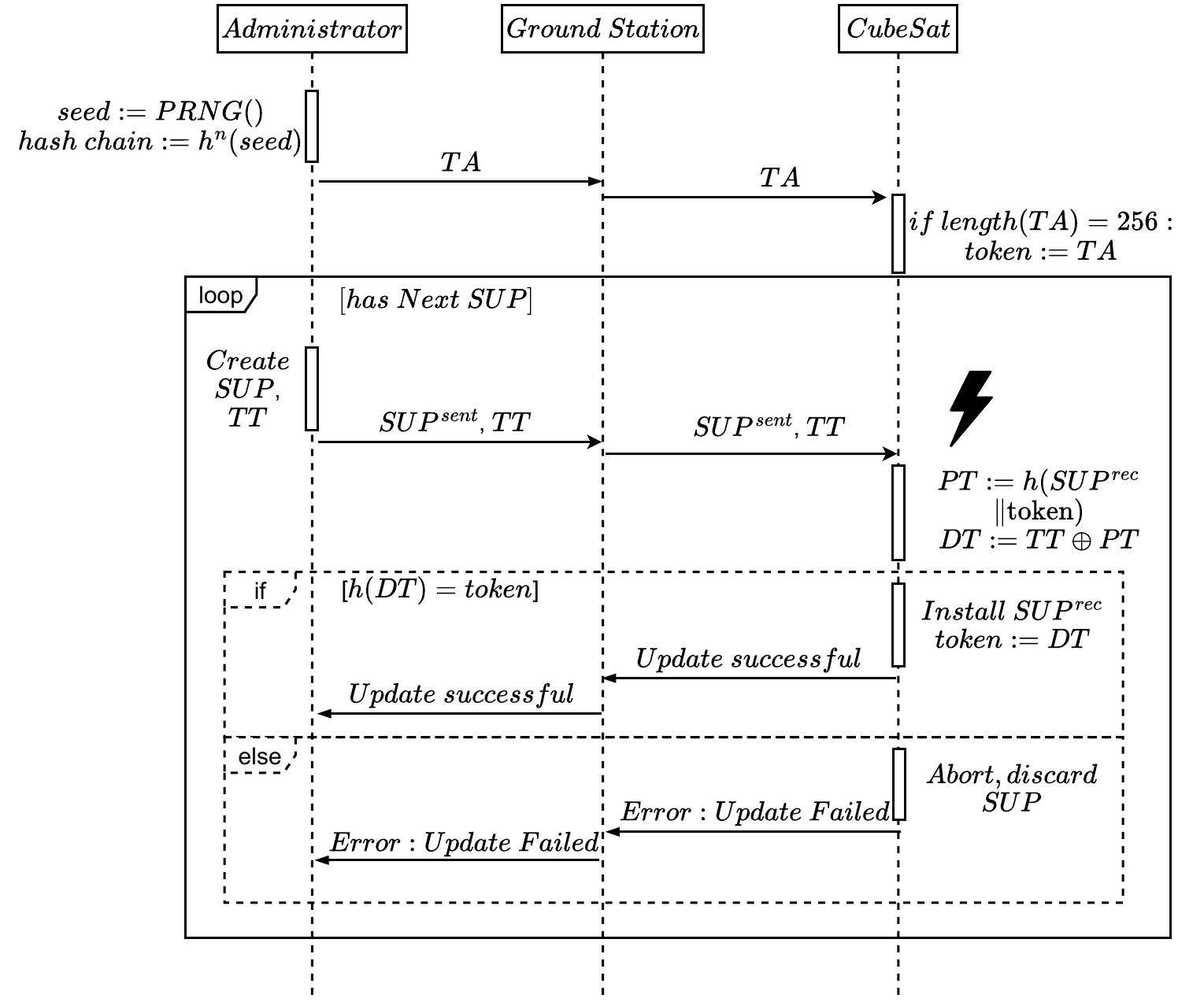}
    \caption{Sequence diagram illustrating CSUM for secure software update.}
    \label{fig:sequence_diagram_hashauth}
\end{figure}

This section explains the tasks executed by the administrator, GS, and CS within the proposed scheme. \figurename~\ref{fig:sequence_diagram_hashauth} presents a sequence diagram that illustrates CSUM for securely installing software updates on CSs.
\subsubsection{Administrator operation}
Several crucial tasks are undertaken to ensure the proper functioning of CSUM at the administrator's end. These tasks include generating a hash chain and creating a $SUP$ with the corresponding $TT$. The administrator initializes the scheme by generating a random seed; the generated seed is hashed iteratively to create a hash chain. Each $SUP$ sent to CS is accompanied by its corresponding $TT$, formulated as shown in Eq.~\eqref{eq:transmission_token}. Algorithm~\ref{algo:TT_admin} presents the pseudocode for $TT$ generation by the administrator for each~$SUP$.
\par
The administrator and GS can communicate in an encrypted or plain text manner. However, we require an integrity mechanism while sending $SUP$s without encryption. We especially require an encrypted connection between the administrator and GS only while sending $TA$ during the initial setup phase. For regular $SUP$s, the administrator creates and sends $SUP$ and the corresponding $TT$ to the GS. Algorithm~\ref{algo:TT_admin} details the psuedocode for this case.

\begin{algorithm}[ht!]
    \caption{Pseudocode for $TT$ generation.}
    \label{algo:TT_admin}
    \begin{flushleft}
    \hspace*{\algorithmicindent} \textbf{Input:} $SUP$, $AT_{curr}$, $AT_{prev}$ \\
    \hspace*{\algorithmicindent} \textbf{Output:} $TT$
    \end{flushleft}
    \begin{algorithmic}[1]       
        \If{$h(AT_{curr}) = AT_{prev}$} \Comment{Validate AT}
            \State $PT = h(SUP \mathbin\Vert AT_{prev})$
            \State $TT := AT_{curr} \oplus PT$
            \State \Return $TT$
        \Else
            \State Invalid $AT_{curr}$ and $AT_{prev}$ combination
        \EndIf
    \end{algorithmic}
\end{algorithm}

\subsubsection{Ground station operations}
GS receives $SUP$ and $TT$ from the administrator. It forwards both $SUP$ and $TT$ to CS in unencrypted form. GS has more resources than CS in terms of computational power. It enables GS to manage communication links with multiple satellites simultaneously and process large amounts of data transmitted from such satellites.

\subsubsection{CubeSat operations}
CS extracts both $SUP$ and its corresponding $TT$ from a transmission received from the GS. Subsequently, CS deciphers $AT_{curr}$ using received $SUP$, $TT$ and $AT_{prev}$, using Eq.~\eqref{eq:token_decipher}.
\begin{align}
    \label{eq:token_decipher}
    AT_{curr} \coloneqq TT \oplus h(SUP \mathbin\Vert AT_{prev})
\end{align}
\par
Following token extraction, CS verifies the authenticity of deciphered $AT_{curr}$ by comparing its hash with $AT_{prev}$ as shown in Eq.~\eqref{eq:auth}. Successful verification confirms the integrity and authenticity of $SUP$, and thus, CS installs $SUP$ and updates $AT_{prev}$. Algorithm~\ref{algo:AT_cubesat} provides psuedocode for AT extraction, AT verification, and $SUP$ installation at CS.
\par~\par~
\begin{algorithm}[H]
    \caption{Pseudocode for $AT$ extraction and verification.}
    \label{algo:AT_cubesat}
    \begin{flushleft}
    \hspace*{\algorithmicindent} \textbf{Input:} $SUP^{rec}$, $TT$ \\
    \hspace*{\algorithmicindent} \textbf{Output:} True/False
    \end{flushleft}
    \begin{algorithmic}[1]       
        \While{$\text{hasNextUpdate()}$}
            \State $PT := h(SUP^{rec} \mathbin\Vert token)$
            \State $DT := TT \oplus PT$
            \If{$h(DT) = token$} \Comment{Validate $DT$}
                \State Install $SUP^{rec}$ on CS
                \State $token := DT$
                \State Send ``Update successful'' message to GS
                \State \Return True
            \Else
                \State Abort update
                \State Send ``Error: Update Failed'' message to GS
                \State \Return False
            \EndIf
        \EndWhile
    \end{algorithmic}
\end{algorithm}

\par
\begin{figure}[H]
    \centering
    \includegraphics[width=0.65\columnwidth]{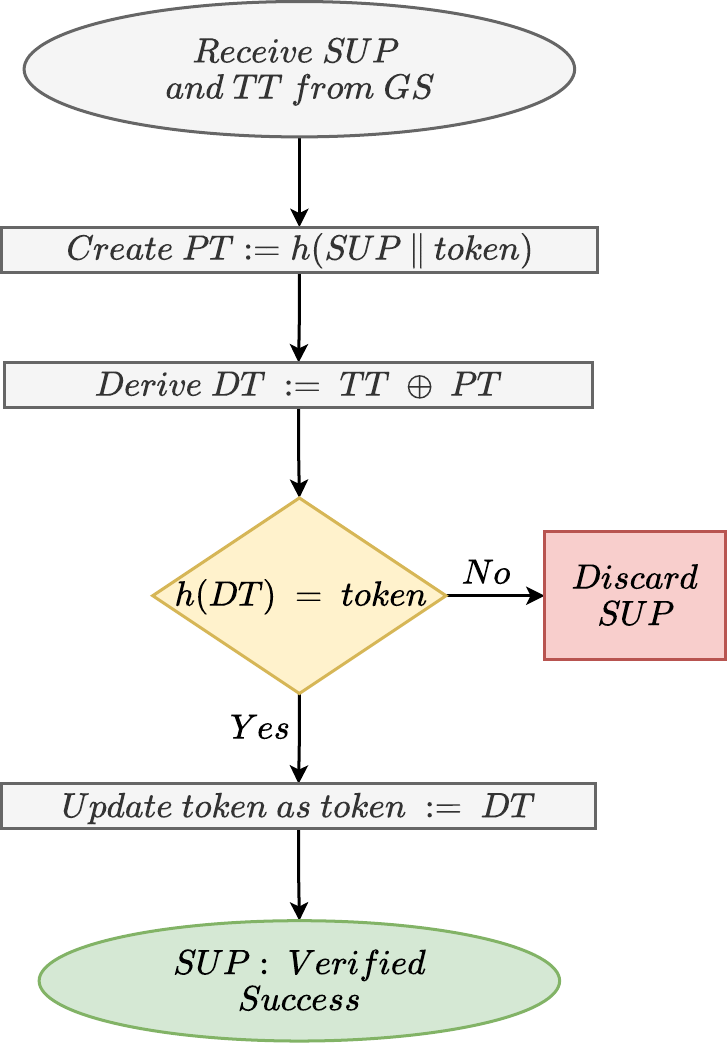}
    \caption{Flowchart describing software verification procedure at CS.}
    \label{fig:sw_verification}
\end{figure}

The flowchart depicted in \figurename~\ref{fig:sw_verification} outlines the verification procedure for $TT$ at CS. If CS receive $SUP^{sent}$ and $TT$ without any modification to $SUP$ or $TT$ during transmission then $SUP^{sent} = SUP^{rec}$, $DT = AT_{curr}$. Here, $token$ is a local variable stored in CS such that other parties are unable to access it.

\subsection{Sample execution} 
\label{subsection:run}
\begin{figure}[ht!]
	\centering
	\includegraphics[width=0.95\columnwidth]{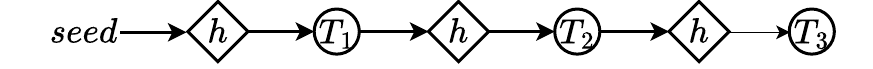}
	\caption{A sample hash chain of length three.}
	\label{fig:example}
\end{figure}

We now present a sample execution of CSUM to illustrate its working better. Consider a hash chain of size three as depicted in \figurename~\ref{fig:example}. After hashing the seed thrice, the administrator obtains $T_3$, which is securely stored on CS in the $token$ variable as its $TA$~(prior to CS's launch in space). Given the hash chain has a length of three, it can support two\footnote{In practice, CSUM can support one more update. However, it would require using the variable length seed. For the sake of engaging only fixed-size $token$s~(to maintain constant network overheads), we suggest using seed-based validation only in extreme scenarios, e.g., wiping CS with a final update or re-initializing the~hash~chain.} successful software updates in CSUM. During such updates, two distinct $SUP$ are broadcast to CS. \tablename~\ref{tab:example} lists the computations at both the administrator and CS for each update iteration.

\begin{table}[htbp]
	\centering
	\caption{CSUM dry run for a hash chain of length three.}
	\label{tab:example}
 \renewcommand{\arraystretch}{1.1}
	\begin{tabular}{|c|c|c|c|}
		\hline
                \multirow{2}{*}{\textbf{Site}} & \multirow{2}{*}{\textbf{Variable}} & \multicolumn{2}{c|}{\textbf{Update iteration}} \\ \cline{3-4} 
                      &                           & \multicolumn{1}{c|}{\textbf{One}}     & \textbf{Two}    \\ \hline
		\multirow{4}{*}{\rotatebox[origin=c]{90}{Administrator~}} & $SUP$       & $sw1$                                   & $sw2$                                   \\ \cline{2-4} 
  & $AT_{prev}$ & $T_3$                             & $T_2$                                \\ \hhline{~|-|-|-|} \cline{2-4}
  & \cellcolor{yellow}$\mathbf{AT_{curr}}$ & \cellcolor{yellow}$\mathbf{T_2}$                & \cellcolor{yellow}$\mathbf{T_1}$                                \\ \hhline{~|-|-|-|} \cline{2-4}
		& $TT$        & 
		$\begin{array}{@{}l@{}}
			TT_1 = T_2 \oplus \\
			\quad \quad \quad h(sw1 \Vert T_3)
		\end{array}$ &
		$\begin{array}{@{}l@{}}
			TT_2 = T_1 \oplus \\
			\quad \quad \quad h(sw2 \Vert T_2)
		\end{array}$ \\  \hline

		\multirow{4}{*}{\rotatebox[origin=c]{90}{CS}}            & $SUP^{rec}$       & $sw1$              & $sw2$   \\ \cline{2-4}
  &    $token$     & $T_3$                & $T_2$                             \\ \cline{2-4} 
		&   $PT$        &      $h(sw1 \Vert T_3)$      &   $h(sw2 \Vert T_2)$  \\ \hhline{~|-|-|-|}
		\cline{2-4} 
		& \cellcolor{yellow}$\mathbf{DT}$        & \cellcolor{yellow}${\mathbf{T_2}}$                                 &\cellcolor{yellow} $\mathbf{T_1}$                                \\  \hhline{~|-|-|-|}\hline
	\end{tabular}
\end{table}
\par
\textit{$1^{st}$ software update:} The administrator prepares $SUP=sw1$ and generates $TT=T_2 \oplus h(sw1 \mathbin\Vert T_3)$ as described in Eq.~\eqref{eq:transmission_token}. Here, $AT_{curr}=T_2$ and $AT_{prev}=T_3$. Upon receiving a software update, CS creates $PT=h(SUP^{rec} \mathbin \Vert token)$ and uses it to derive $DT=TT \oplus PT$ as specified in Eq.~\eqref{eq:token_decipher}. If $SUP^{rec}$ is equal to $sw1$, then $DT$ resolves to $T_2$~(matching $AT_{curr}$). With $DT=T_2$ and $token=T_3$, CS checks whether $h(DT)=token$ to authenticate $SUP^{rec}$~(cf. Eq.~\eqref{eq:auth}). Given the relationship between $T_2$ and $T_3$~(cf. \figurename~\ref{fig:example}), the verification is successful. CS now installs $sw_1$ and updates its $token=T_2$.
\par
\textit{$2^{nd}$ software update:} The administrator prepares $SUP=sw2$ and calculates $TT=T_1 \oplus h(sw2 \mathbin \Vert T_2)$. Now, $AT_{curr}=T_1$ and $AT_{prev}=T_2$.  Upon receiving a software update, CS calculate $PT=h(SUP^{rec} \mathbin \Vert token)$ and use it to derive $DT=TT \oplus PT$. If $SUP^{rec}$ is equal to $sw2$, then $DT$ resolves to $T_1$~(matching $AT_{curr}$). With $DT=T_1$ and $token=T_2$, CS checks whether $h(DT)=token$ to authenticate $SUP^{rec}$. Given the relationship between $T_1$ and $T_2$, the verification is successful. CS now installs $sw_2$ and updates its $token=T_1$.
\par
To summarize, CSUM sends $AT_{curr}$ to CS in such a way that only CS that has the knowledge of corresponding $AT_{prev}$ can extract it. 

\section{Experimental Evaluation}
\label{section:evaluation}
We performed experiments on a Lenovo ThinkPad P14s Gen 2 machine. We implemented our proposed scheme using python 3.7 and utilized four common open-source packages including Putty, Notepad++, FileZilla, and Audacity with sizes $<$ 16~MB to simulate $SUP$ being sent to CS.  
\par
Encryption/decryption-based schemes are traditionally employed to provide authentication, confidentiality, and integrity. However, they may not inherently guarantee data freshness, which is crucial to prevent replay attacks. Moreover, decryption operations at CS can impose a significant computational burden. Signature-based schemes primarily focus on ensuring authentication and integrity with the potential to provide freshness depending on the specific implementation. However, signature verification operations at CS can also introduce computational overhead. 
\par
\begin{figure}[ht!]
    \centering
    \includegraphics[width=\columnwidth]{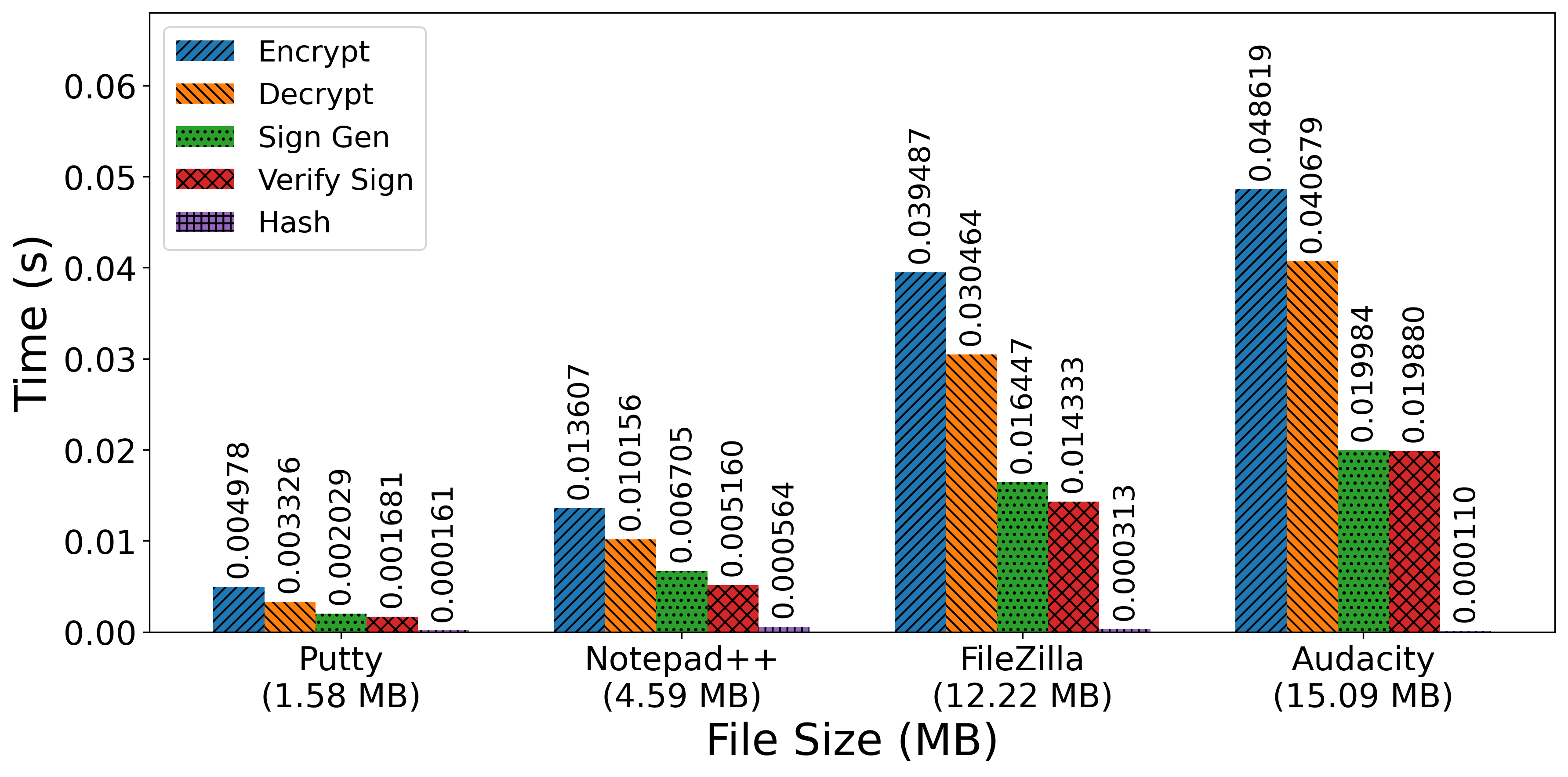}
    \caption{Timing information for various software of cryptographic primitives.}
    \label{fig:timing}
\end{figure}

To this end, we designed an experiment with different commonly used software applications, including Putty~v0.80~\cite{putty}, NotePad++~v8.6~\cite{notepad++}, FileZilla~v3.66.4~\cite{filezilla}, and Audacity~v3.4.2~\cite{audacity}. We performed the following cryptographic operations: encryption, decryption, signature generation, signature verification, and hashing on these open-source software packages to collect timing data. We utilize RSA with a 2048-bit key with PSS padding~\cite{bellare1998pss} for signing, AES with a 256-bit key in CBC mode for encryption, and SHA-256 for hashing. We used \textit{hazmat} layer from the Python cryptography library~\cite{py_cryptography} to implement cryptographic primitives. The results are shown in \figurename~\ref{fig:timing}. Encryption, signature generation, and hashing operations are performed at administrator~(without any resource limitation) while decryption, signature verification, and hashing operation are done at CS~(with limited resources).
\par
As shown in \figurename~\ref{fig:timing}, we observe that hashing operations are consistently the fastest across various file sizes, taking only 0.000161s for a 1.58~MB file, 0.000564s for a 4.59~MB file, 0.000313s for a 12.22~MB file, and 0.000110s for a 15.09~MB file. Signature verification, a process that ensures data integrity and authenticity, is more time-consuming than hashing, with times ranging from 0.001681s for the smallest file to 0.019880s for the largest file. Decryption times, which ensures confidentiality, data integrity and authenticity, is even more expensive then signature verification, requiring 0.003326s for the smallest file and scaling up to 0.040679s for the largest tested file.
\par
We have taken previous literature and our results into account and decided to utilize hash-based mechanism to design CSUM. CSUM takes a different approach by employing a hash-based mechanism, which inherently assures authentication, integrity, and freshness while using a hash chain. Freshness is assured by the sequential nature of the hash chain, which prevents the reuse of old hashes. CSUM simplifies the verification process by requiring only a single hash operation at CS, significantly reducing computational overhead.

\textit{Performance:} \tablename~\ref{tab:hashchain_performance} shows the performance analysis of the hash chain, particularly, the time taken to generate the hash chain and the verification time. CSUM takes less than 0.01s to generate a hash chain for 10,000 updates, and it gradually increases to 0.056s for 50,000 updates. This rise indicates a proportional increment in the computational load associated with a larger number of targeted updates. Naturally, the number of targeted updates directly affects to number of times our system has to hash the seed to create the hash chain. On another side, verifying 10,000 updates requires about 0.18s, which increases to 0.805s for 50,000 updates. This increase is steeper since more operations are involved in verifying $SUP$s than in generating a hash chain. Overall, both the hash chain generation time and verification of $SUP$s grow linearly with the number of $SUP$s.

\begin{table}[ht]
\centering
\caption{Performance analysis of hash chains.}
\begin{tabular}{ccc}
\hline
\begin{tabular}[c]{@{}c@{}}Number of\\$SUP$s\end{tabular} &
  \begin{tabular}[c]{@{}c@{}}Time taken to \\ generate hash \\ chain (s)\end{tabular} &
  \begin{tabular}[c]{@{}c@{}}Verification by our\\ approach for every \\ single $SUP$ in chain (s)\end{tabular} \\
  \hline
10,000 & \textless 0.01   & 0.179558 \\
20,000 & 0.019748 & 0.316285 \\
30,000 & 0.031258 & 0.509595 \\
40,000 & 0.040098 & 0.635965 \\
50,000 & 0.055507 & 0.805479 \\ 
\hline
\label{tab:hashchain_performance}
\end{tabular}
\end{table}
\par
\textit{Network overhead:}
The network overhead in our approach is independent of the size of a $SUP$, and it is determined by the fixed output length of the utilized hash function, \ie 256~bits in our implementation. Therefore, our proposed scheme has a constant network overhead.

\section{Security Analysis}
\label{section:security_analysis}
\begin{figure}[H]
	\centering
	\includegraphics[width=\columnwidth]{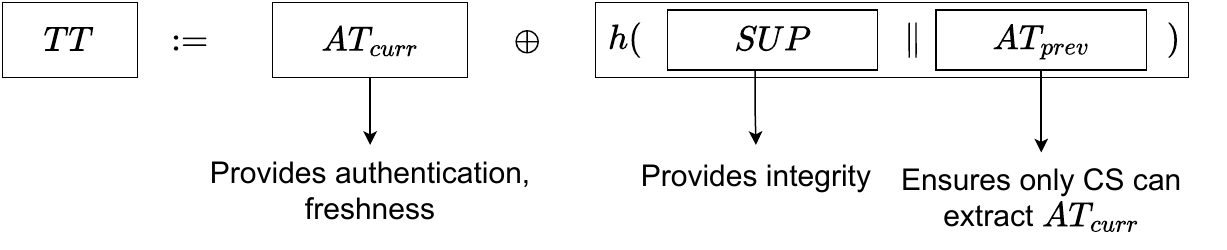}
	\caption{$TT$ structure: authentication with $AT_{curr}$, integrity via $SUP$ hash, and freshness with hash chain continuity.}
	\label{fig:security_equation}
\end{figure}

\figurename~\ref{fig:security_equation} outlines security features provided by different components of CSUM's $TT$. In particular, $TT$ integrates authentication with $AT_{curr}$, ensures the integrity of $SUP$ through hashing, and maintains freshness with hash chain continuity. This approach effectively safeguards against unauthorized access and data manipulation. To summarize, CSUM leverages a hash chain for generating $AT$ and embeds $AT$ within $TT$ to preserve authenticity, integrity, and freshness for each $SUP$ transmission.
\subsection{Authentication}
Authentication is facilitated by utilizing $AT_{curr}$, which is known only to the legitimate entities involved in the software update process. In CSUM, each $AT_{curr}$ is cryptographically linked to $AT_{prev}$ as shown in Eq.~\eqref{eq:auth}, preventing adversaries from determining the pre-image of $AT_{prev}$ in polynomial time without breaking pre-image resistance property of a cryptographic hash function. Including $AT_{curr}$ within $TT$ enables CS to verify the origin of $SUP$, preventing impersonation and unauthorized access.
\par
Successfully guessing $AT_{curr}$ allows the recovery of $AT_{prev}$, enabling an adversary to create a valid $TT$ for their altered update. However, the likelihood of accurately guessing $AT_{curr}$ is extremely low, i.e.,  $1/2^{256}$. Alternatively, correctly guessing $AT_{prev}$ permits the extraction of $AT_{curr}$, not by deriving pre-image but by employing Eq.~\eqref{eq:token_decipher}. However, the probability of guessing $AT_{prev}$ remains similarly low.

\subsection{Integrity}
The integrity of $SUP$ is maintained by incorporating the hash of $SUP$ concatenated with $AT_{prev}$ in a $TT$, rendering $TT$ update specific. Any changes in $SUP$ data would result in a distinct hash output, generating different $TT$ and an invalid $AT_{curr}$ as depicted in Eq.~\eqref{eq:auth}. If Eq.~\eqref{eq:auth} is not satisfied during verification at CS, it indicates tampering with $SUP$, leading to rejection. Now we delve into a brief proof-by-contradiction for asserting ``\textit{Any unauthorized modification to $SUP$ is detectable.}''
\begin{itemize}
	\item Assume an adversary can make an undetectable modification to $SUP$.
	\item An undetectable modification implies that after altering $SUP$, the resulting $TT$ is still valid, \ie $TT  \coloneqq AT_{curr} \oplus h(SUP' \mathbin\Vert AT_{prev})$.
	\item This requires $h(SUP \mathbin\Vert AT_{prev}) = h(SUP' \mathbin\Vert AT_{prev})$. However, finding such $SUP$$'$, where $h(SUP \mathbin\Vert AT_{prev})$ is equal to $h(SUP' \mathbin\Vert AT_{prev})$ breaks second pre-image resistance of a cryptographic hash function. 
\end{itemize}
Our assumption leads to a contradiction, establishing that integrity in CSUM is maintained via detection of modifications.

\subsection{Freshness}
Freshness is ensured by leveraging the sequential nature of the hash chain used to generate $AT$. Since each $AT_{curr}$ is used only once and is replaced by the next in the hash chain for the subsequent update, ensuring that each $TT$ is unique. We are utilizing each token in the hash chain as a one-time password. CS verifies Eq.~\eqref{eq:auth} to ascertain the freshness of an update, mitigating replay attacks where adversaries attempt to resend old $SUP$s with previously valid $TT$s.
\par
\textit{Attack prevention:} CSUM inherently protects against common attacks including: 
\begin{enumerate}
	\item \textit{Replay attacks:} CSUM utilizes one-time tokens derived from a hash chain, making it resistant to replay attacks. CS rejects previously captured $SUP$ along with their respective $TT$s as they fail to correct $AT_{curr}$, leading to the automatic rejection of the update.
	\item \textit{Malicious update attacks:} CSUM generates a portion of the $TT$ using a hash of $SUP$ combined with $AT_{prev}$ to tackle malicious update attacks. Any modification in $SUP$ renders the resulting $TT$ invalid, preventing CS from deriving the correct $AT_{curr}$ and resulting in the rejection of a malicious update.
	\item \textit{Token swapping attacks:} CSUM is resistant to token swapping attacks as CS could not derive correct $AT_{curr}$ from old authentic TT, leading to the automatic rejection of the outdated update. 
\end{enumerate}

\section{Limitations} 
\label{section:discussion}
The potential limitations of CSUM are as follows:
\begin{enumerate}
    \item \textit{Unique hash chain for each satellite:} CSUM requires a unique hash chain for every satellite. However, it may not be an issue as the administrator and GS are resource-rich. The size of the hash chain, which the administrator generates, is minimal. For instance, storing a hash chain of length 100 requires only 3.2~KB. Moreover, the size of each token is the same as the size of the hash function output~(making it suitable for the administrator, GS, and even CS). Finally, there are a finite number of satellites under the control of GS.
    \item \textit{Re-initialize hash chain:} CSUM requires re-initializing the hash chain once all the tokens are used in the existing hash chain. Authors in~\cite{goyal2004re, zhao2005improved, zhang2008novel} demonstrate efficient ways to re-initialize the hash chain, which can be extended to our work as well.
    \item \textit{Lack of confidentiality:} An increasing number of CS projects are moving towards open-source software~\cite{shalashov2021review}. The emphasis has shifted from the principle of \textit{security through obscurity} towards a model that prefers transparency and collaborative security practices. Open-source projects benefit from this paradigm by allowing broader inspection and collective improvement. Therefore, the security of CSUM does not rely on concealing the software's components but rather on the robustness of the cryptographic methods implemented.
\end{enumerate}
	
\section{Conclusion and future work}
\label{section:conclusion}
The fundamental requirements of authentication, integrity, and data freshness are essential for ensuring the security of $SUP$s broadcast to CS. In this paper, we propose a lightweight scheme that ensures authentication, integrity, and data freshness for CS software updates, providing a practical solution specifically tailored to the resource-constrained environment of CS. We validate the practical feasibility and efficacy of our proposed approach by developing a proof of concept. CSUM does have limitations, such as the requirement for unique hash chains for each CS and the need for hash chain re-initialization. Despite these challenges, our work significantly improves the overall security and the performance of the software update procedure for CS.
\par
In the future, we will explore a scalable and secure group update scheme for a cluster of CSs, eliminating the need for unique $AT$ per satellite/update.
\balance
\bibliographystyle{IEEEtran}
\balance
\bibliography{bib}
\balance
\end{document}